\documentclass[aps,prl,twocolumn,numerical,superscriptaddress,nofootinbib,showpacs,showkeys,longbibliography]{revtex4-1}

\usepackage{graphicx,color}
\usepackage{amsmath,amssymb,amsfonts}

\newcommand{\be}{\begin{equation}}
\newcommand{\ee}{\end{equation}}
\newcommand{\ba}{\begin{eqnarray}}
\newcommand{\ea}{\end{eqnarray}}
\newcommand{\ban}{\begin{eqnarray*}}
\newcommand{\ean}{\end{eqnarray*}}

\def\v2{\mbox{$v_2$}}

\def\sqrtsNN{\mbox{$\sqrt{s_{\mathrm{NN}}}$}}

\begin{document}

\title{Quantification of the Chiral Magnetic Effect in Au+Au collisions at $\sqrt{s_{\mathrm{NN}}}=200$~GeV}
\medskip

\author{Roy~A.~Lacey} 
\email{Roy.Lacey@stonybrook.edu}
\affiliation{Depts. of Chemistry \& Physics, Stony Brook University, Stony Brook, New York 11794, USA}

\author{Niseem~Magdy} 
\email{niseemm@gmail.com}
\affiliation{Department of Physics, University of Illinois at Chicago, Chicago, Illinois 60607, USA}

%

%
%
%
%

\date{\today}

\begin{abstract}
The Multi-Phase Transport model, AMPT, and the Anomalous Viscous Fluid Dynamics model, AVFD,  are used to assess a possible chiral-magnetically-driven charge separation ($\Delta S$) recently measured with the ${R_{\Psi_2}(\Delta S)}$ correlator in Au+Au collisions at $\sqrtsNN~=~200$ GeV. The Comparison of the experimental and simulated ${R_{\Psi_2}(\Delta S)}$ distributions indicates that background-driven charge separation is insufficient to account for the measurements. The  AVFD model calculations, which explicitly account for CME-driven anomalous transport in the presence of background, indicate a CME signal quantified by the $P$-odd Fourier dipole coefficient ${a_1'}\approx 0.5\%$ in mid-central collisions. 
A similar evaluation for the $\Delta\gamma$ correlator suggests that only a small fraction of this signal ($f_{\rm CME}=\Delta\gamma_{\rm CME}/\Delta\gamma \approx 25\%$) is measurable with this correlator in the same collisions. The related prediction for signal detection in isobaric collisions of Ru+Ru and Zr+Zr are also presented.
%
\end{abstract}

\pacs{25.75.-q, 25.75.Gz, 25.75.Ld}
\maketitle

Heavy-ion collisions at the Relativistic Heavy Ion Collider (RHIC) and the Large Hadron Collider (LHC) 
lead to the production of a magnetized chiral relativistic quark-gluon 
plasma (QGP) \cite{Kharzeev:2004ey,Liao:2014ava,Miransky:2015ava,Huang:2015oca,Kharzeev:2015znc}, 
akin to the primordial plasma produced in the early Universe \cite{Rogachevskii:2017uyc,Gorbunov:2011zz} and 
several degenerate forms of matter found in compact stars \cite{Weber:2004kj}. Pseudo-relativistic analogs 
include Dirac and Weyl semimetals \cite{Vafek:2013mpa,Burkov:2015hba,Gorbar:2017lnp}. 
 The study of anomalous transport in the QGP can give fundamental insight not only 
on the complex interplay of chiral symmetry restoration, axial anomaly and gluon 
topology\cite{Moore:2010jd,Mace:2016svc,Liao:2010nv,Kharzeev:2015znc,Skokov:2016yrj}, but also on 
the evolution of magnetic fields in the early Universe \cite{Joyce:1997uy,Tashiro:2012mf}. 

A major anomalous process predicted to occur in the magnetized QGP is the 
chiral magnetic effect (CME) \cite{Fukushima:2008xe}.
It is characterized  by the vector current:
\begin{equation}
\vec{J}_V = \frac{N_{c}e\vec{B}}{2\pi^2}\mu_A, {\rm for}\, \mu_A\neq 0,
\end{equation}
where $N_c$ is the color factor,  $\vec{B}$ is the magnetic field and $\mu_A$ is the axial chemical potential 
that quantifies the axial charge asymmetry or imbalance between right- and left-handed quarks in the 
plasma \cite{Fukushima:2008xe,Son:2009tf,Zakharov:2012vv,Fukushima:2012vr}. 
Experimentally, the CME manifests as the separation of electrical charges along the 
$\vec{B}$-field~\cite{Kharzeev:2004ey,Fukushima:2008xe}. This stems from the fact that 
the CME preferentially drives charged particles, 
originating from the same ``P-odd domain'', along or opposite to the $\vec{B}$-field 
depending on their charge. 

The charge separation can be quantified via measurements of the first $P$-odd 
sine term ${a_{1}}$, in the Fourier decomposition of the charged-particle azimuthal 
distribution~\cite{Voloshin:2004vk}:
\begin{eqnarray}\label{eq:a1}
{\frac{dN^{\rm ch}}{d\phi} \propto 1 + 2\sum_{n} (v_{n} \cos(n \Delta\phi) + a_n \sin(n \Delta\phi)  + ...)}\
\end{eqnarray}
where $\mathrm{\Delta\phi = \phi -\Psi_{RP}}$ gives the particle azimuthal angle
with respect to the reaction plane (${\rm RP}$) angle, and ${v_{n}}$ and ${a_{n}}$ denote the
coefficients of the $P$-even and $P$-odd Fourier terms, respectively. 
A direct measurement of the P-odd coefficients  $a_1$, is not possible due to the 
strict global $\cal{P}$ and $\cal{CP}$ symmetry of QCD.
However, their fluctuation and/or variance $\tilde{a}_1= \left<a_1^2 \right>^{1/2}$ can 
be measured with charge-sensitive  correlators such as the 
$\gamma$-correlator~\cite{Voloshin:2004vk}  and the ${R_{\Psi_m}(\Delta S)}$ 
correlator~\cite{Magdy:2017yje,Magdy:2018lwk,Huang:2019vfy,Magdy:2020wiu}.

The $\gamma$-correlator measures charge separation as:
\begin{eqnarray}
\gamma_{\alpha\beta} =& \left\langle \cos\big(\phi_\alpha +
\phi_\beta -2 \Psi_{\rm EP}\big) \right\rangle, \nonumber \quad
\Delta\gamma =& \gamma_{\rm OS} - \gamma_{\rm SS},
\label{eq:2}
\end{eqnarray}
where $\Psi_{\rm EP}$ is the azimuthal angle of the event plane, $\phi$ denote the particle azimuthal emission angles, 
$\alpha,\beta$ denote the electric charge $(+)$ or $(-)$
 and SS and OS represent same-sign ($++,\,--$) and opposite-sign ($+\,-$) charges.

The $R_{\Psi_m}(\Delta S)$ correlator~\cite{Magdy:2017yje,Magdy:2018lwk}
is constructed for the $m^{\rm th}$-order event plane $\Psi_m$, as the ratio:
\be
R_{\Psi_m}(\Delta S) = C_{\Psi_m}(\Delta S)/C_{\Psi_m}^{\perp}(\Delta S), \, m=2,3 ,
\label{eq:4}
\ee
where $C_{\Psi_m}(\Delta S)$ and $C_{\Psi_m}^{\perp}(\Delta S)$ are correlation functions
that quantify charge separation $\Delta S$, parallel and perpendicular (respectively) to 
the $\vec{B}$-field. 
$C_{\Psi_2}(\Delta S)$ measures both CME- and background-driven charge separation 
while $C_{\Psi_2}^{\perp}(\Delta S)$ measures only background-driven charge separation. 
The absence of a strong correlation between the orientation of the $\Psi_3$ plane 
and the $\vec{B}$-field, also renders $C_{\Psi_3}(\Delta S)$ and $C_{\Psi_3}^{\perp}(\Delta S)$
insensitive to a CME-driven charge separation, but not to the background, so it can give  
additional insight  into the relative importance of background-driven and CME-driven 
charge separation~\cite{Magdy:2017yje,Magdy:2018lwk}.  

Recently, the STAR collaboration reported new charge separation measurements for the ${R_{\Psi_2}(\Delta S)}$ correlator, which suggested a possible  CME-driven charge separation in Au+Au collisions \cite{1799840}. Here, we use the AMPT \cite{Lin:2004en}  and AVFD~\cite{Shi:2017cpu,Jiang:2016wve} models with varying degrees of charge separation, characterized by the dipole term ${a_1'}$, to calibrate the correlator and extract an estimate of the magnitude of the suggested CME signal. We also use the AVFD model to calibrate the correlators for isobaric collisions (Ru+Ru and Zr+Zr) and estimate the magnitude of a possible signal for each isobar and the signal and background differences between them. Both models are known to give good representations of the experimentally measured particle yields, spectra, flow, etc.,\cite{Lin:2004en,Ma:2016fve,Ma:2013gga,Ma:2013uqa,Bzdak:2014dia,Proceedings:2017uei}. Thus, they include realistic estimates for several backgrounds such as flow and flow fluctuations, resonance decays, local charge conservation, and global momentum conservation. They also provide an important benchmark for evaluating the interplay between possible CME- and background-driven charge separation in actual data.

Anomalous transport due to the CME is explicitly implemented in the AVFD model. An in-depth account of this implementation can be found in Refs.~\cite{Shi:2017cpu} and \cite{Jiang:2016wve}. In brief, the second-generation Event-by-Event  version of the model, called E-by-E AVFD,  uses Monte Carlo Glauber initial conditions to simulate the evolution of fermion currents in the QGP, in concert with the bulk fluid evolution implemented in the VISHNU hydrodynamic code, followed by a URQMD hadron cascade stage. A time-dependent magnetic field $B(\tau) = \frac{B_0}{1+\left(\tau / \tau_B\right)^2}$, acting in concert with a nonzero initial axial charge density $n_5/s$, is used to generate a CME current (embedded in the fluid dynamical equations), leading to a charge separation along the magnetic field. The peak values $B_0$, obtained from event-by-event simulations~\cite{Bloczynski:2012en}, are used with a relatively conservative lifetime $\tau_B=0.6$ fm/c. The commonly used estimate based on the strong chromo-electromagnetic fields in the early-stage glasma is adopted for the initial axial charge density arising from gluonic topological charge fluctuations. The anomalous transport in AVFD results in primordial charge separations on the freeze-out surface and final-state charge separation of the produced hadrons, quantified by the $P$-odd Fourier coefficients ${a_1’}$ and ${\tilde{a}_{1}}$ respectively. Note that  ${\tilde{a}_{1}} < {a_1’}$ due to smearing and dilution from resonance feed-down in the hadronic phase. The signal loss, relative to ${a_1’}$, is about 30\%  for Au+Au collisions \cite{Shi:2017cpu}.

Anomalous transport from the CME is not implemented in AMPT. Instead, modifications have been made to the model to mimic CME-induced charge separation in the partonic phase~\cite{Ma:2011uma}. This is accomplished by switching the $p_y$ values of a fraction of the downward moving $u$ ($\bar{d}$) quarks with those of the upward moving $\bar{u}$ ($d$) quarks to produce a net charge-dipole separation. Here, the $x$ axis is along the direction of the impact parameter $b$, the $z$ axis points along the beam direction, and the $y$ axis is perpendicular to the $x-z$ plane, {\em i.e}, the direction of the proxy $\vec{B}$-field. The strength of this proxy ``primordial'' charge separation is regulated by the fraction $f_{p}$ \cite{Ma:2011uma,Huang:2019vfy}:
\begin{equation}
f_{p} = \frac{N_{\uparrow(\downarrow)}^{+(-)}-N_{\downarrow(\uparrow)}^{+(-)}}{N_{\uparrow(\downarrow)}^{+(-)}+N_{\downarrow(\uparrow)}^{+(-)}},
\quad f_{p} =\frac{4}{\pi}{a'}_{1}
 \label{eq-f}
\end{equation}
where $N$ is the number of a given species of quarks, $``+''$ and $``-''$ denote positive and negative charges, respectively, 
and $\uparrow$ and $\downarrow$ represent the directions along and opposite to that of the $y$ axis. 

The fraction $f_p$, is related to the $P$-odd dipole coefficient for the produced hadrons $\tilde{a}_{1}$ (cf. Eqs.~\ref{eq:a1} and \ref{eq-f}). However, the ``partonic'' dipole coefficient ${a'}_{1}$ has a  different relationship to the final hadrons'  $\tilde{a}_1$, than that for AVFD and other models~\cite{Jiang:2016wve,Shi:2017cpu,Sun:2018idn}. A comparison of the models for Au+Au collisions at $\sqrtsNN = 200$~GeV indicates that $\tilde{a}_{1}(\rm AMPT)$ and $\tilde{a}_{1}(\rm AVFD)$ are approximately linearly related, and the events for both models give the same $R_{\Psi_2}(\Delta S)$ correlator response for ${a_1'}(\rm AMPT) \approx 6\times {a_1'}(\rm AVFD)$. This factor reflects the fact that the signal loss in AMPT, due to parton cascade and resonance feed-down \cite{Ma:2011uma}, is much larger than that for AVFD. Because CME-driven anomalous transport is explicit in the AVFD model, we use ${a_1'}(\rm AVFD)$, $n_5/s$ and $\tilde{a}_{1}(\rm AVFD)$ as quantitative measures in the following.

Simulated AMPT and AVFD events, generated for a broad set of $a_1^{'}$ values, were analyzed with both the $\Delta\gamma$ and the $R_{\Psi_2}(\Delta S)$ correlators to facilitate the calibrations necessary for estimating the magnitude of a possible CME-driven signal from the measurements. Event selections and cuts included charged particles with $|\eta| < 1.0$ and transverse momentum $0.2 < p_T < 2$~GeV/$c$. To enhance the statistical significance of the AMPT events, the $\Psi_{2}$ plane was determined with charged hadrons in the range $2.5< \eta <4.0$. The charge separation of charged hadrons for $|\eta| < 1.0$ was then measured relative to $\Psi_{2}$.
%
%
\begin{figure}[t]
\includegraphics[width=1.0\linewidth, angle=-00]{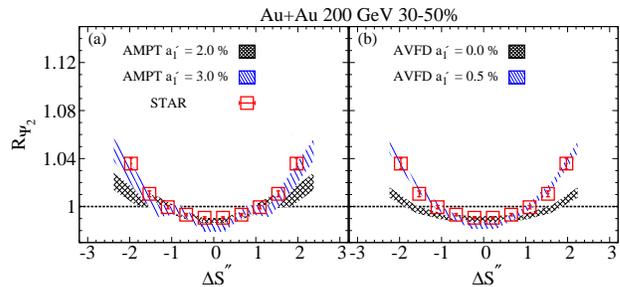}
\caption{ Comparison of the measured and simulated $R_{\Psi_2}(\Delta S)$ correlators for 
 30-50\% Au+Au collisions at $\sqrtsNN = 200$~GeV. The AMPT and AVFD model comparisons are 
shown for several  values of the ${a_1'}(\rm AMPT)$ and ${a_1'}(\rm AVFD)$ dipole coefficient as indicated. 
}
\label{fig1} 
\end{figure} 

Figure~\ref{fig1} shows a representative comparison between the experimental $R_{\Psi_2}(\Delta S^{''})$  distribution (open squares) and those simulated with AMPT (a) and AVFD (b) events for $30-50\%$ central  Au+Au collisions. In these plots, the charge separation is scaled ($\Delta S^{''}$) to account for the effects of particle-number fluctuations and the event-plane resolution~\cite{Magdy:2017yje}. The magnitude of the charge separation is encoded in the variance $\sigma^2_{R_{\Psi_{2}}}$ (or width $\sigma_{R_{\Psi_{2}}}$) of the concave-shaped $R_{\Psi_2}(\Delta S^{''})$ distribution~\cite{Magdy:2017yje,Magdy:2018lwk,Magdy:2020wiu}. 

The simulated distributions for ${a_1'}(\rm AMPT) \approx 3.0\%$ (panel a) and ${a_1'}(\rm AVFD) \approx  0.50\%$ (panel b) show good agreement with the data. By contrast, the distributions simulated for ${a_1'} = 0.0\%$ (background only) do not agree with the data, as shown in Fig.~\ref{fig1} (b). This disagreement indicates that background-driven charge separation alone is insufficient to account for the measurement.
Here, it is noteworthy that the background-driven charge separation for AVFD is constrained by ensuring good agreement between the experimental and simulated $R_{\Psi_{2}}(\Delta S^{''})$  distributions for $N_{\rm chg} \approx 20$ or mean centrality $\sim 75\%$.  For such collisions, background-driven charge separation predominates over CME-driven charge separation due to the approximately random $\vec{B}$-field orientations relative to the $\Psi_{\rm 2}$ event plane. The ${a_1'}$-independent $R_{\Psi_{2}}(\Delta S^{''})$ distributions obtained with AVFD for the same centrality selection, confirms this expected insensitivity to the signal in peripheral collisions.
%
%
\begin{figure}[tb]
\includegraphics[width=1.0\linewidth, angle=-00]{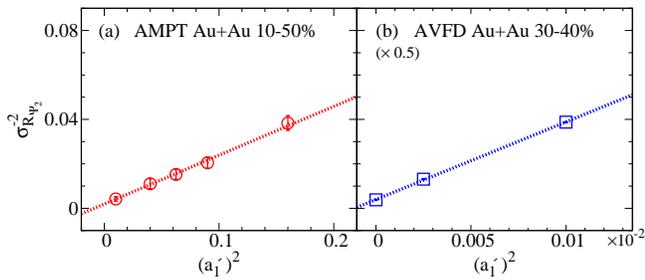}
\vskip -0.10in
\caption{ Inverse variances $\mathrm{\sigma^{-2}_{R_{\Psi_2}}}$ extracted from the 
$R_{\Psi_2}(\Delta S^{''})$ correlator  vs. the $P$-odd dipole coefficient $({a_1'})^2$. 
Results are shown for 10-50\% central AMPT events (a) and 30-40\% central 
AVFD events for Au+Au collisions at $\sqrtsNN = 200$~GeV; the dotted lines represent the lines of best fit.
} 
\label{fig2} 
\end{figure} 

To further calibrate the signal strength, we extracted the variance $\mathrm{\sigma^2_{R_{\Psi_2}}}$ of the $R_{\Psi_{2}}(\Delta S^{''})$ distributions, obtained for several values of ${a_1'}{(\rm AMPT)}$ and ${a_1'}{(\rm AVFD)}$ in Au+Au collisions; the ${a_1'}{(\rm AVFD)}$ values correspond to the $n_5/s$ values of 0.0, 0.1 and 0.2 respectively.  Figs.~\ref{fig2} (a) and (b) show representative plots of the inverse variance $\mathrm{\sigma^{-2}_{R_{\Psi_2}}}$ vs. $({a_1'})^2$ for 10--50\% central AMPT events (a) and 30--40\% central AVFD events (b). Both calibration curves indicate an essentially linear dependence of $\mathrm{\sigma^{-2}_{R_{\Psi_2}}}$ on $({a_1'})^2$ (note the dotted lines of best fit), albeit with slope differences that reflect the observation that ${a_1'}(\rm AMPT) \approx 6\times {a_1'}(\rm AVFD)$ for similar response at the same centrality. The relatively small associated intercepts in Figs.~\ref{fig2} (a) and (b), indicate an influence from background-driven charge separation, possibly dominated by the effects of local charge conservation.

The comparison of the Au+Au data to the AVFD calibration curve in Fig.~\ref{fig2} (b) indicates the value  ${a_1'} = 0.50\pm 0.03$\% for mid-central collisions. This value, which corresponds to $n_5/s=0.1$, is similar to the estimate obtained from the comparison between the experimental and simulated $R_{\Psi_{2}}(\Delta S^{''})$ distributions (cf.  Fig.~\ref{fig1} (b)). The inverse variances $\mathrm{\sigma^{-2}_{R_{\Psi_2}}}$ for ${a_1'} = 0.0$\% and ${a_1'} = 0.50$\% also allow evaluation of the fraction $f_{\rm CME}$ of ${\sigma^{-2}_{R_{\Psi_2}}}(0.50\%)$ attributable to the CME as:
\[
f_{\rm CME}(R_{\Psi_{2}}) =\frac{[{\sigma^{-2}_{R_{\Psi_2}}}(0.50\%)-{\sigma^{-2}_{R_{\Psi_2}}}(0.0\%)]}
{[{\sigma^{-2}_{R_{\Psi_2}}}(0.50\%)]} \approx 76\%.
\]
This sizable $f_{\rm CME}(R_{\Psi_{2}})$ value is a good benchmark for the sensitivity of the $R_{\Psi_{2}}(\Delta S^{''})$ correlator to 
CME-driven charge separation in these collisions.
 
The AMPT and AVFD events used to calibrate the $R_{\Psi_{2}}(\Delta S^{''})$ correlator were also employed to calibrate the $\Delta\gamma$ correlator and predict the magnitude of the CME-driven signal strength expected from its measurements. The calibration procedure involving the analysis of AMPT and AVFD events with varying degrees of the signal is similar to that for the $R_{\Psi_{2}}(\Delta S^{''})$ correlator.
%
%
\begin{figure}[t]
\includegraphics[width=1.0\linewidth, angle=-00]{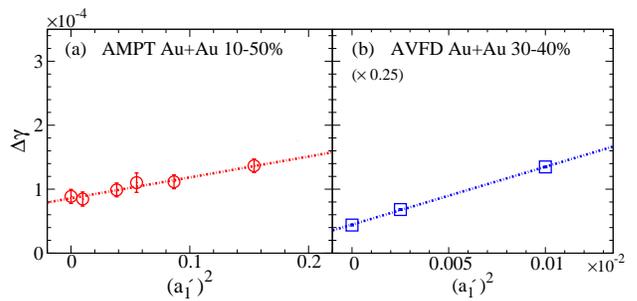}
\vskip -0.10in
\caption{ ${\Delta\gamma}$ vs. $({a_1'})^2$.  Results are shown for 10-50\% central AMPT events (a) and 30-40\% central 
AVFD events (b) for Au+Au collisions at $\sqrtsNN = 200$~GeV; the dotted lines represent the lines of best fit.
} 
\label{fig3} 
\end{figure} 

Figures \ref{fig3} (a) and (b) show the resulting calibration curves [$\Delta\gamma$ vs. $({a_1'})^2$)] obtained for AMPT and AVFD events, respectively. Both plots indicate the expected linear dependence of $\Delta\gamma$ on $({a_1'})^2$ [note the dotted lines of best fit]. However, the non-negligible intercepts indicate significant background contributions to the $\Delta\gamma$ values for $({a_1'})^2 > 0$. The slopes of the calibration curves also reflect the signal losses \cite{Ma:2011uma,Huang:2019vfy} alluded to earlier. The steeper slope for AVFD events, apparent in Fig.~\ref{fig3} (b), is also in line with the hierarchy of more significant signal losses for AMPT events than for AVFD events.

For the value ${a_1'}(\rm AVFD) = 0.5\%$ [extracted with the $R_{\Psi_{2}}(\Delta S^{''})$ correlator] and its AMPT equivalent, Figs.~\ref{fig3} (a) and (b)  indicate that the fraction of the $\Delta\gamma$ value attributable to the CME, 
	\[f_{\rm CME}(\Delta\gamma)=\frac{[{\Delta\gamma}(0.50\%) - {\Delta\gamma}(0.0\%)]}
{[{\Delta\gamma}(0.50\%)]} \approx 25\%.\]
 This $f_{\rm CME}(\Delta\gamma)$ value is consistent with the recent measurements reported in Ref.~\cite{Adam:2020zsu}, albeit with sizable uncertainties. It is also roughly a factor of three times smaller than $f_{\rm CME}(R_{\Psi_{2}})$, suggesting that the $R_{\Psi_2}(\Delta S)$ correlator is more sensitive for this signal level \cite{Sun:2018idn,Huang:2019vfy,Shi:2019wzi}.

The AVFD model was also used to calibrate the $R_{\Psi_{2}}(\Delta S^{''})$ and $\Delta\gamma$ correlators and predict the magnitude of the CME-driven signal strength expected in isobaric collisions of Ru+Ru and Zr+Zr at $\sqrtsNN = 200$~GeV. The calibration, which followed the procedure outlined earlier, involved the analysis of 30-40\% central AVFD events for the same $n_5/s$ values employed in the Au+Au simulations {\em i.e.}, $n_5/s= 0, 0.1$ and $0.2$ respectively.

\begin{figure}[t]
\includegraphics[width=1.1\linewidth, angle=-00]{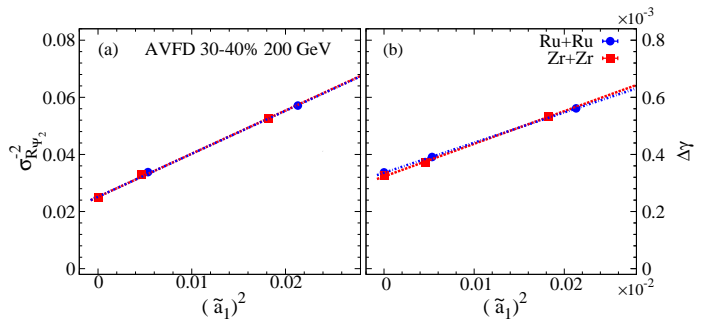}
\vskip -0.10in
\caption{ Comparison of the $R_{\Psi_2}(\Delta S^{''})$ (a) and ${\Delta\gamma}(\Psi_{2})$ (b) calibration curves for Ru+Ru and Zr+Zr collisions at $\sqrtsNN = 200$~GeV. Results are shown for the 30-40\% centrality selection; the dotted lines represent the lines of best fit.
} 
\label{fig4} 
\end{figure}

Figures \ref{fig4} (a) and (b) show the respective calibration curves for the isobaric collisions. Both plots indicate the expected linear dependence of ${\sigma^{-2}_{R_{\Psi_2}}}$ and $\Delta\gamma$ on $(\tilde{a}_{1})^2$ with signal differences [between the isobars] that depend on the magnitude of $\tilde{a}_{1}$. The plotted values for the latter correspond to $n_5/s = 0, 0.1$ and $0.2$ respectively. The non-negligible intercepts, also apparent in the figures, indicate significant background contributions to both ${\sigma^{-2}_{R_{\Psi_2}}}$ and $\Delta\gamma$ for $(\tilde{a}_{1})^{2} > 0.0$. These contributions are reflected in the values $f_{\rm CME}(R_{\Psi_{2}}) \approx 25\%$ and $f_{\rm CME}(\Delta\gamma) \approx 13\%$ evaluated for the isobars for $n_5/s= 0.1$. They indicate that the sensitivity of both correlators is significantly reduced compared to that for Au+Au collisions simulated for $n_5/s =0.1$, albeit with an approximate factor of two difference between $f_{\rm CME}(R_{\Psi_{2}})$ and $f_{\rm CME}(\Delta\gamma)$. This difference suggests that, for the 30-40\% isobaric collisions, background-driven charge separation already begins to prevail over CME-driven charge separation. More central collisions might be needed to achieve better sensitivity. Note that the background dominates in peripheral Au+Au collisions but not in mid-central collisions.

The isobaric signal difference is shown in Figs.~\ref{fig4} (a) and (b);  for $n_5/s= 0.1$ it is much smaller than the respective signal magnitude for each isobar, relative to the background, and will require substantial statistical significance to measure. Therefore, CME characterization in these collisions will benefit significantly from the planned measurements of the respective signal magnitude for each isobar, in addition to measurements of the isobaric signal and background differences.
%
%

In summary, AVFD model simulations that incorporate varying degrees of CME- and background-driven charge separation are used to quantify a possible chiral-magnetically-driven charge separation measured with the ${R_{\Psi_2}(\Delta S)}$ correlator in Au+Au collisions at $\sqrtsNN~=~200$ GeV. The simulations which quantify the CME  via the $P$-odd Fourier dipole coefficient ${a_1'}$ indicate the value ${a_1'} = 0.50\pm 0.03$\% in mid-central collisions, consistent with a  modest CME signal.  A similar calibration for the $\Delta\gamma$ correlator suggests that, only a small fraction of this signal ($f_{\rm CME}=\Delta\gamma_{\rm CME}/\Delta\gamma \approx 25\%$) is measurable with the $\Delta\gamma$ correlator in the same collisions. A further calibration for isobaric collisions of Ru+Ru and Zr+Zr, suggests that CME characterization in these collisions not only require measurement of the isobaric signal difference, but also the respective signal magnitude for each isobar and an estimate of the  background difference between them.


\section*{Acknowledgments}
\begin{acknowledgments}
This research is supported by the US Department of Energy, Office of Science, Office of Nuclear Physics, 
under contracts DE-FG02-87ER40331.A008  (RL) and DE-FG02-94ER40865 (NM). 
%
\end{acknowledgments}
%
%
\bibliography{lpvpub} 
\end{document}